%% file: main-cav.tex
\documentclass{llncs}
\usepackage[ruled,linesnumbered,noend]{algorithm2e}
\usepackage{citesort}
\usepackage{multicol}
\usepackage{amssymb}
\usepackage{amsmath,bbm}
\usepackage{latexsym,epic,eepic}
\usepackage{complexity}
\usepackage{multirow}
\usepackage{algorithmic}
\usepackage{url}
\usepackage{btran}
\usepackage{times}
\usepackage{paralist}
\usepackage{tikz}
\usetikzlibrary{shapes}
\usetikzlibrary{automata,positioning}

\newenvironment{to-do}
{ \rule{1ex}{1ex}\hspace{\stretch{1}} \bfseries}
{ \hspace{\stretch{1}}\rule{1ex}{1ex} \vspace{1ex}}

\include{comm-anthony}


\newif\ifdraft\draftfalse
\ifdraft
\newcommand\anthony[1]{{\color{blue}
[#1 - \textbf{Anthony}]}}
\newcommand\philipp[1]{{\color{red}
[#1 - \textbf{Philipp}]}}
\newcommand\chihduo[1]{{\color{brown}
[#1 - \textbf{Chihduo}]}}

\else
\newcommand\todo[1]{}
\newcommand\anthony[1]{}
\newcommand\philipp[1]{}
\newcommand\chihduo[1]{}
\newcommand\sunjun[1]{}

\fi

\newcommand\shortlong[2]{#1}

\usepackage{color}
\shortlong{}{\pagestyle{plain}}

\title{Probabilistic Bisimulation for Parameterized Systems (Technical Report)\thanks{
	This research was sponsored in part by the ERC Starting Grant
    759969 (AV-SMP), ERC Synergy project 610150 (ImPACT), the 
	DFG project 389792660-TRR 248 ({Perspicuous Computing}),
      the Swedish Research Council (VR)
    under grant~2018-04727, and by the Swedish Foundation for Strategic
    Research (SSF) under the project WebSec (Ref.\ RIT17-0011).
}}

\institute{}
\author{
    Chih-Duo Hong\inst{1} \and
    Anthony W. Lin\inst{2} \and
    Rupak Majumdar\inst{3} \and
    Philipp R\"ummer\inst{2}
}
\institute{
    Oxford University, United Kingdom \and
    TU Kaiserslautern, Germany \and
    Uppsala University, Sweden \and
    Max Planck Institute for Software Systems, Germany
}
\date{}

\begin{document}

\maketitle

\begin{abstract}
    \input{abstract}
\end{abstract}

\input{introduction}
\input{prelim}
\input{systems}

\input{model}
\input{model-words}

\input{case-study}
\input{learning}

\bibliographystyle{abbrv}
\bibliography{references}

%
\input{appendix}

\end{document}

%% file: comm-anthony.tex
\newcommand{\OMIT}[1]{}
\newcommand{\ACT}{\ensuremath{\text{\sf ACT}}}

\newcommand{\N}{{\ensuremath{\tt N}} }

\newcommand{\Nat}{{\ensuremath{\mathbb{N}}}}

\newcommand{\defn}[1]{\textit{#1}} 
\newcommand{\mcl}[1]{\ensuremath{\mathcal{#1}}}
\newcommand{\NatInt}[2]{\ensuremath{\{#1,\ldots,#2\}}}
\renewcommand{\mod}[2]{\ensuremath{#1~{\rm mod}~#2}}
\newcommand{\sem}[1]{[\![#1 ]\!]}

\newcommand{\zerodist}{\ensuremath{\overline{0}}}

\newcommand{\transysDom}{S}
\newcommand{\Kripke}{\ensuremath{\langle S; \{\to_a\}_{a \in \ACT}\rangle}}
\newcommand{\pref}{\preceq}
\newcommand{\zero}{{\sf zero}}
\newcommand{\one}{{\sf one}}
\newcommand{\toss}{{\sf toss}}
\newcommand{\tail}{{\sf tail}}
\newcommand{\head}{{\sf head}}
\newcommand{\reset}{{\sf end}}
\newcommand{\prefSucc}{\prec_{succ}}
\newcommand{\eqlen}{\text{eqL}}

\newcommand{\struct}{\mathfrak{S}}

\newcommand{\univ}{\,\mathfrak{U}\,}
\newcommand{\univDesc}{\langle \ialphabet^*; \pref, \eqlen, \{l_a\}_{a \in
\ialphabet}\rangle}
\newcommand{\Tree}{\ensuremath{\text{\scshape Tree}}}

\newcommand{\transys}{\ensuremath{\langle S; \to\rangle}}

\newcommand{\transysMDP}{\ensuremath{\langle S; \{\delta_a \}_{a\in\ACT}\rangle}}
\newcommand{\transysPTS}{\transysMDP}

\newcommand{\Rec}{\ensuremath{\text{\scshape Rec}}}

\newcommand{\empseq}{\ensuremath{\varepsilon}}

\newcommand{\ialphabet}{\ensuremath{\Sigma}}

\newcommand{\Real}{\ensuremath{\mathbb{R}}}

\newcommand{\problemx}[3]{
\par\noindent\underline{\sc#1}\par\nobreak\vskip.2\baselineskip
\begingroup\clubpenalty10000\widowpenalty10000
\setbox0\hbox{\bf Instance: }\setbox1\hbox{\bf Question: }
\dimen0=\wd0\ifnum\wd1>\dimen0\dimen0=\wd1\fi
\vskip-\parskip\noindent
\hbox to\dimen0{\box0\hfil}\hangindent\dimen0\hangafter1\ignorespaces#2\par
\vskip-\parskip\noindent
\hbox to\dimen0{\box1\hfil}\hangindent\dimen0\hangafter1\ignorespaces#3\par
\endgroup}

\newcommand{\Paths}{\ensuremath{\text{\scshape Paths}}}

\newcommand{\Aut}{\ensuremath{\mathcal{A}}} 
\newcommand{\transrel}{\ensuremath{\Delta}} 
\newcommand{\tran}[1]{\ensuremath{\stackrel{#1}{\longrightarrow}}}
\newcommand{\controls}{\ensuremath{Q}} 
\newcommand{\finals}{\ensuremath{F}} 
\newcommand{\Lang}{\ensuremath{\mathcal{L}}} 
\newcommand{\AutRun}{\ensuremath{\rho}} 

\newcommand{\arity}{\ensuremath{\text{\texttt{ar}}}}

\newcommand{\rarw}{\rightarrow}

\newcommand{\ModelRun}{\ensuremath{\pi}}

\newcommand{\Size}[1]{\ensuremath{\|#1\|}}


%% file: abstract.tex
Probabilistic bisimulation is a fundamental notion of process equivalence for
probabilistic systems. It has important applications, including 
the formalisation of the anonymity property of several communication protocols. 
While there is a large body of work on verifying probabilistic bisimulation for finite systems, the problem is in general undecidable
 for parameterized systems, i.e., for infinite families
of finite systems with an arbitrary number $n$ of processes. 
In this paper we provide a general framework for reasoning about
probabilistic bisimulation for parameterized systems. Our approach is in
the spirit of software verification, wherein we encode proof rules for 
probabilistic bisimulation and use a decidable first-order theory to specify 
systems and candidate bisimulation relations, which can then be checked 
automatically against the proof rules.

We work in the framework of regular model checking, and 
specify an infinite-state system as a regular relation described by a 
first-order formula over a universal automatic structure, i.e., a logical theory
over the string domain.
For probabilistic systems, we show how probability values (as well as the
required operations) can be encoded naturally in the logic.
Our main result is that one can specify 
the verification condition of whether a 
given regular binary relation is a probabilistic bisimulation
as a regular relation.
Since the first-order theory of the universal automatic structure
is decidable, we obtain an effective method for verifying 
probabilistic bisimulation for infinite-state systems, given a regular 
relation as a candidate proof. 
As a case study, we show that our framework is sufficiently expressive for
proving the anonymity property of the parameterized
dining cryptographers protocol 
and the parameterized grades protocol. 
Both of these protocols hitherto could not be verified by existing automatic methods. 
Moreover, with the help of standard automata learning algorithms, we show that the
candidate relations can be synthesized fully automatically, making the verification
fully automated. 
\OMIT{
Given a candidate binary relation specified as a formula in this theory,
we show that it is decidable to check whether it is a probabilistic
bisimulation of two given automatic transition systems (i.e. systems that can be
specified in the theory), under the assumption that they are of bounded 
branching. 
}
\OMIT{
As case studies, we show that our framework is sufficiently expressive to 
capture both the sequential and concurrent version of Parameterized 
Dining Cryptographers Protocol (i.e., with any number of cryptographers), whose 
anonymity property hitherto could not be automatically proven by existing 
automated verification methods. 
}
\OMIT{
We propose a framework for specifying probabilistic infinite 

Privacy is an important property in many communication protocols, which ensures 
that 
certain information remains a secret throughout the course of the protocol. 
Anonymity is an example of such a property, which requires that
an observer cannot identify the agents participating a protocol 
from the actions they execute.
}

%% file: introduction.tex
\section{Introduction}
\label{sec:intro}

Equivalence checking using bisimulation relations plays a fundamental 
role in formal verification. 
Bisimulation is the basis for substitutability of systems: if two systems are bisimilar,
their behaviors are the same and they satisfy the same formulas in
expressive temporal logics.
The notion of bisimulation is defined both for deterministic \cite{Milner89}
and for probabilistic transition systems \cite{LS91}. 
In both contexts, checking bisimulation has many applications, such as proving correctness of
anonymous communication protocols \cite{CNP09}, reasoning about knowledge \cite{halpern-book},
program optimization 
\cite{KTL09}, and optimizations for computational problems 
(e.g. language equivalence and minimization) of finite 
automata \cite{BP13}. 

The problem of checking bisimilarity of two given systems has been widely studied. 
It is decidable in polynomial-time for both probabilistic and non-probabilistic \emph{finite-state} systems
\cite{Baier96,DHS03,VF10,CBW12}.
These algorithms form the basis of practical tools for checking bisimulation.
For infinite-state systems, such as parameterized versions of communication
protocols (i.e. infinite families of finite-state systems with an arbitrary
number $n$ of processes), the problem is undecidable in general.
Most research hitherto has focused on
identifying decidable subcases (e.g. strong
bisimulations for pushdown systems for probabilistic and non-probabilistic cases
\cite{S05,FJKW12,roadmap}), rather than on providing tool support for practical problems. 


In this paper, we propose a first-order verification approach---inspired by
software verification techniques---for reasoning about bisimilarity for infinite-state
systems.
In our approach, we provide first-order logic \emph{proof rules} to determine
if a given binary relation is a bisimulation.
To this end, we must find an \emph{encoding} of systems and relations
and a \emph{decidable first-order theory} that can 
formalize the system, the property, and the proof rules. 
We propose to use the decidable first-order theory of the \emph{universal 
automatic structure} \cite{BLSS03,BG04}.
Informally, the domain of the theory is a set of words over a finite alphabet $\ialphabet$,
and it captures the first-order theory of 
the infinite $|\ialphabet|$-ary tree with a relation that relates strings
of the same level.
The theory can express precisely the class of all \emph{regular 
relations} \cite{BLSS03} (a.k.a.  automatic relations \cite{BG04}), which
are relations $\varphi(x_1,\ldots,x_k)$ over strings $\ialphabet^*$ that
can be recognized by synchronous multi-tape automata.
It is also sufficiently powerful to capture 
many classes of non-probabilistic infinite-state systems and regular model 
checking \cite{anthony-thesis,TL08,TL10,rmc-survey,RMC}.

\OMIT{
We demonstrate the effectiveness of the approach by 
verifying challenging examples of parameterized systems from the literature, 
including proving the anonymity property of the parameterized dining 
Cryptographers Protocol \cite{chaum1988dining} and of the parameterized grades 
protocol \cite{apex},
both of which were only automated for some fixed parameters using finite-state
model checkers or equivalence checkers (e.g. see \cite{apex,prism}).}

We demonstrate the effectiveness of the approach by encoding and
automatically verifying some challenging examples from the literature of parameterized systems in our logic:
the anonymity property of the parameterized dining cryptographers protocol~\cite{chaum1988dining}
and the grades protocol \cite{apex}. 
These examples were only automatically verified for some fixed parameters using 
finite-state
model checkers or equivalence checkers (e.g. see \cite{apex,prism}).
Just as invariant verification for software separates out the proof rules 
(verification conditions in a decidable logic)
from the synthesis of invariants,
we separate out proof rules for bisimulation from the
synthesis of bisimulation relations.
We demonstrate how recent developments in
generating and refining candidate proofs as automata
(e.g.~\cite{LR16,Vardhan-thesis,HV05,CHLR17,Neider13,NT16,GNMR16,LMN16})
can be used to automate the search of proofs,
making our verification fully ``push button.'' 

%

\smallskip
\noindent\textbf{Contributions.} 
Our contributions are as follows. 
First, we show how probabilistic infinite-state 
systems can be faithfully encoded in the first-order theory of universal 
automatic 
structure. In the past, the theory has been used to reason about qualitative
liveness of weakly-finite MDPs (e.g. see \cite{LR16,fairness}), which allows the
authors to
disregard the actual non-zero probability values. To the best of our knowledge,
no encoding of probabilistic transition systems in the theory was available.
In order to be able to effectively encode probabilistic systems, our theory
should typically be two-sorted: one sort for encoding the configurations, and 
the other for encoding the probability values. We show how both sorts (and the
operations required for the sorts) can be encoded in the universal automatic
structure, which requires only the domain of strings. In the sequel, such
transition systems will be called \emph{regular transition systems}.

Second, using the \emph{minimal deviation assumption}~\cite{LS91},
i.e., there exists $\varepsilon > 0$
such that all transition probabilities of the system are multiples of $\varepsilon$,
we show how the verification condition of whether a given regular binary relation is
a probabilistic bisimulation can be encoded in the theory. The decidability of the
first-order theory over the universal automatic structure gives us an 
effective means of checking probabilistic bisimulation for regular transition
systems. 
In fact, the theory can be easily reduced to the weak monadic theory
WS1S of one successor (therefore, allowing highly optimized tools like Mona 
\cite{monasecrets} and Gaston \cite{fiedor_lazy_2017})
by interpreting finite words as finite sets~\cite{rubin-thesis,CL07}.

Our framework requires the encoding of the systems and the proofs in the
first-order theory of the universal automatic structure. Which interesting
examples can it capture? Our third contribution is to provide
two examples from the literature of parameterized verification:
the anonymity property of the parameterized dining cryptographers protocol
\cite{chaum1988dining} and of the parameterized grades protocol \cite{apex}.
We study two versions of dining cryptographers protocol in this paper:
the classical version where the secrets are single bits, and
a generalized version where the secrets are bit-vectors
of arbitrary length. 

Thus far, our framework requires a candidate proof to be supplied by the user.
Our final contribution is to demonstrate how standard techniques from the
synthesis literature (e.g. automata learning 
\cite{LR16,Vardhan-thesis,HV05,CHLR17,Neider13,NT16,GNMR16,LMN16})
can be used to fully automate the proof search.
Using automata learning, we successfully pinpoint regular proofs
for the anonymity property of the three protocols:
the two dining cryptographers protocols are verified in 6 and 28 seconds, respectively,
and the grades protocol  in 35 seconds.

\smallskip
\noindent\textbf{Other related work.} 
The verification framework we use in this paper can be construed as a regular 
model checking \cite{rmc-survey} framework using regular relations. The
framework uses first-order logic as the language, which makes it convenient to
express many verification conditions (as is well-known from first-order theorem
proving \cite{bradley-book}). The use of the universal automatic structure
allows us to express two different sorts (configurations and probability values)
in one sort (i.e. strings). Most work in regular model checking focuses on 
safety and liveness properties (e.g. 
\cite{rmc-survey,rmc-thesis,RMC,ltl-mso,BLW03,Neider13,HV05,fairness,anthony-thesis,TL10,LR16,Vardhan-thesis}).

Some automated techniques can prove the anonymity property 
of the dining cryptographers protocol and the grades protocol
in the finite case, e.g., the PRISM model
checker~\cite{prism,prism-dcp} and language equivalence by the tool APEX 
\cite{apex}. To the best of our knowledge, our method is the first automated 
technique proving the anonymity property of the protocols in the parameterized case. 

Our work is in spirit of deductive software verification
(e.g., \cite{ESCJava,Shoham,ivy,bradley-book,DBLP:series/lncs/10001,DBLP:conf/lpar/Leino10}),
where one provides inductive invariants manually, and a tool
automatically checks correctness of the candidate invariants. 
In theory, our result yields a fully-automatic procedure by
enumerating all candidate regular proofs, and at the same time enumerating
all candidate counterexamples (note that we avoid undecidability by restricting
attention to proofs encodeable as regular relations).
In our implementation, we use recent advances in automata-learning based synthesis to efficiently
encode the search \cite{LR16,CHLR17}.

\OMIT{
We also mention the extremely successful verification framework of constrained 
horn clauses \cite{horn-clause}, which fits better with SMT (which typically
does not handle quantified formulas). Checking 

Mention horn clause thing and say that you cannot
model bisimulation using horn clauses (or in datalog). 
}

%% file: prelim.tex
\section{Preliminaries}
\label{sec:prelim}

\noindent
\textbf{General notation.}
We use $\Nat$ to denote non-negative integers.
Given $a,b \in \Real$, we use a standard notation
$[a,b] := \{ c\in \Real : a \le c \le b \}$ to denote real intervals.
Given a set $S$, we use $S^*$ to denote the set of all finite
sequences of elements from $S$. The set $S^*$ always includes the empty
sequence which we denote by $\empseq$. 
We call a function $f : S \to [0,1]$ a \emph{probability distribution over $S$}
if $\sum_{s\in S} f(s) = 1$. 
We shall use $I_s$ to denote the probability distribution $f$ with $f(s)=1$,
and $\mathcal{D}_S$ to denote the set of probability distributions over $S$.
Given a function $f : X_1\times\cdots\times X_n\to Y$,
the \emph{graph} of $f$ is the relation 
$\{(x_1,...,x_n,f(x_1,...,x_n)) : \forall i\in \NatInt{1}{n}.\ x_i \in X_i\}$.
Whenever  a relation $R$ is an equivalence relation over set $S$, we use $S/R$
to denote the set of equivalence classes created by $R$.
Depending on the context, we may use $p\,R\,q$ or $R(p,q)$ to denote $(p,q)\in R$.
\OMIT{
    Given two sets of words
    $S_1, S_2$, we use $S_1\cdot S_2$ to denote the set $\{ v\cdot w: v\in S_1,
    w\in S_2\}$ of words formed by concatenating words from $S_1$ with words from
    $S_2$. Given two relations $R_1,R_2 \subseteq S \times S$, we define
    their composition as $R_1 \circ R_2 := \{ (s_1,s_3) : \exists s_2.\ (s_1,s_2)
    \in R_1 \wedge (s_2,s_3) \in R_2\}$.}

\smallskip
\noindent
\textbf{Words and automata.}
We assume basic familiarity with word automata.
Fix a finite alphabet $\Sigma$. For each finite word $w := w_1\ldots w_n \in
\Sigma^*$, we
write $w[i,j]$, where $1 \leq i \leq j \leq n$, to denote the segment
$w_i\ldots w_j$. Given an automaton $\mcl{A} := (\Sigma,Q,\delta,q_0,F)$,
a run of $\mcl{A}$ on $w$ is a function $\rho: \NatInt{0}{n}
\rarw Q$ with $\rho(0) = q_0$ that obeys the transition relation $\delta$. 
We may also denote the run $\rho$ by the word $\rho(0)\cdots \rho(n)$ over
the alphabet $Q$. 
The run $\rho$ is said to be \defn{accepting} if $\rho(n) \in F$, in which
case we say that the word $w$ is \defn{accepted} by $\mcl{A}$. The language
$L(\mcl{A})$ of $\mcl{A}$ is the set of words in $\Sigma^*$ accepted by
$\mcl{A}$. 

\smallskip 
\noindent
\textbf{Transition systems.}
We fix a set $\ACT$ of \defn{action symbols}.
A \defn{transition system} over $\ACT$ is a tuple $\struct := \Kripke$,
where $S$ is a set of \defn{configurations}
and $\to_a\ \subseteq S \times S$ is a binary relation over $S$.
We use $\to$ to denote the relation $\bigcup_{a \in \ACT} \to_a$. 
We say that a sequence $s_1 \to \cdots \to s_{n+1}$ is a \defn{path} in $\struct$
if $s_1,...,s_{n+1}\in S$ and $s_i \to s_{i+1}$ for $i\in \NatInt{1}{n}$.
A transition system is called \defn{bounded branching} if the number of configurations
reachable from a configuration in one step is bounded. Formally, this means that
there exists an \emph{a priori} integer $N$ such that for all $s\in\transysDom$, 
$|\{s'\in\transysDom : s\to s' \}| \le N$.

\smallskip
\noindent
\textbf{Probabilistic transition systems.}
A \defn{probabilistic transition system} (\defn{PTS})~\cite{LS91} 
is a structure $\struct := \transysPTS$ where $\transysDom$ is
a set of configurations and $\delta_a:S \to \mathcal{D}_S \cup \{\zerodist \}$
maps each configuration to either a probability distribution or a zero function $\zerodist$.
Here $\delta_a(s)=\zerodist$ simply means that $s$ is a ``dead end'' for action $a$.
We shall use $\delta_a(s, s')$ to denote $\delta_a(s)(s').$
In this paper, we always assume that $\delta_a(s,s')$ is a rational number
and $|\{s' : \delta_a(s, s') \neq 0\}| < \infty$.
The \defn{underlying transition graph} of a~PTS
is a transition system $\langle S; \{\to_a\}_{a\in\ACT}\rangle$ 
such that $s \to_a s'$ iff $\delta_a(s,s')\neq 0$.

In~\cite{LS91}, Larsen and Skou introduced a restriction on PTSs
called \emph{the minimal deviation assumption}, i.e., 
all the transition probabilities appearing are multiples of some $\epsilon>0$.
This assumption is practically sensible since it is satisfied 
by most PTSs we encounter in practice
(e.g.~finite PTSs, probabilistic pushdown automata \cite{EE04}, and 
probabilistic parameterized systems \cite{LR16,fairness}
including our case studies in Section \ref{sec:application}).
The minimal deviation assumption, among others, 
implies that the PTS is bounded-branching
(i.e.~that its underlying transition graph is bounded-branching).
In the sequel, we shall impose this assumption on the PTSs we are dealing with.
\OMIT{
\smallskip
\noindent
\textbf{Bisimulations.}
Let $\struct := \langle S; \{\to_a\}_{a \in \ACT}\rangle$ 
be a transition system.
A \defn{bisimulation} for $\struct$ is a binary relation $R\subseteq S\times S$
such that for all $(p, q) \in R$ and $a\in \ACT$,
\begin{itemize}
    \item If $p \to_a p'$ then there is some $q'\in S$ such that $q \to_a q'$ and $(p',q')\in R$.
    \item  If $q \to_a q'$ then there is some $p'\in S$ such that $p \to_a p'$ and $(p',q')\in R$.
\end{itemize}
We write $p \sim q$ and say $p$ is \defn{bisimilar} to $q$
if and only if $(p,q)\in R$ for some bisimulation $R$ over $S$. 
Note that $\sim$ is itself a bisimulation
and is called the \defn{bisimilarity}.
Furthermore, $\sim$ forms an equivalence relation over $S$
and the notion of bisimulation equivalence
can be specified with
$$
p \sim q \iff \forall a\in \ACT.\ \forall S' \in \transysDom/\!\!\sim\!.\ (p \to_a S' \Leftrightarrow q \to_a S'),
$$
where $p \to_a S'$ means that $p \to_a p'$ for some $p'\in S'$.
}

\smallskip
\noindent
\textbf{Probabilistic bisimulations.}
Let $\struct := \transysPTS$ be a~PTS.
We write $s \tran{\rho}_a S'$ if $\,\sum_{s'\in S'} \delta_a(s,s')=\rho$.
A \defn{probabilistic bisimulation} for $\struct$
is an equivalence relation $R$ over $\transysDom$,
such that $(p,q)\in R$ implies
\begin{equation}\label{defn:pb-proof-rule}
\forall a\in \ACT.\ \forall S' \in S/R.\ (p \tran{\rho}_a S' \Leftrightarrow q \tran{\rho}_a S').
\end{equation}
We say that $p$ and $q$ are \defn{probabilistic bisimilar}~(written as $p \sim q$)
if there is a probabilistic bisimulation $R$ such that $(p,q)\in R$.
A probabilistic bisimulation between two PTSs
$\struct := \transysPTS$ and 
$\struct' := \langle S'; \{\delta_a' \}_{a\in\ACT}\rangle$
is a probabilistic bisimulation for
the disjoint union of $\struct$ and $\struct'$, which is defined as
$\struct\sqcup\struct' := \langle S \uplus S'; \{\delta_a'' \}_{a\in\ACT}\rangle$
where 
$\delta_{a}''(s) := \delta_{a}(s)$ for $s\in S$,
and  
$\delta_{a}''(s) := \delta_{a}'(s)$ for $s\in S'$.
We say $\struct$ and $\struct'$ are \emph{probabilistic bisimilar}
if there is a probabilistic bisimulation $R \supseteq \mathcal I \times \mathcal I'$
for $\struct\sqcup\struct'$,
such that $\mathcal I$ and $\mathcal I'$ are the set of initial configurations
of $\struct$ and $\struct'$, respectively.

%% file: systems.tex
\section{Framework of Regular Relations}
\label{sec:logic}

In this section we describe the framework of regular relations for specifying
probabilistic infinite-state systems, properties to verify, and proofs, all in
a uniform symbolic way. The
framework is amenable to automata-theoretic algorithms in the spirit of
\emph{regular model checking} \cite{rmc-survey,RMC}. 

The framework of \emph{regular relations} \cite{BLSS03} (a.k.a. 
\emph{automatic relations} \cite{Bl99}) uses the first-order theory of 
universal\footnote{Here, ``universal'' simply means that all automatic
structures are first-order interpretable in this structure.}
automatic structure 
\begin{eqnarray}\label{defn:univ-auto-struct}
    \univ := \univDesc,
\end{eqnarray}
where $\ialphabet$ is some finite alphabet, $\pref$ is the (non-strict) prefix-of 
relation, $\eqlen$ is the binary equal length predicate, and $l_a$ is a unary
predicate asserting that the last letter of the word is $a$. 
The domain of the structure is the set of finite words over $\Sigma$,
and for words $w,w'\in \Sigma^*$, we have
$w \pref w'$ if{}f there is some $w''\in\Sigma^*$ such that $w\cdot w'' = w'$,
$\eqlen(w,w')$ iff $|w| = |w'|$, and $l_a(w)$ iff there is some $w''\in\Sigma^*$
such that $w = w''\cdot a$.

Next, we discuss
the expressive power of first-order formulas over the universal automatic 
structures, and decision procedures for satisfiability of such formulas.
In Section \ref{sec:model}, we shall describe: (1) how to specify a PTS as a 
first-order formula in $\univ$, and (2) how to 
specify the verification condition for probabilistic bisimulation property 
in this theory. In Section \ref{sec:application}, we shall show that 
the theory is sufficiently powerful for capturing probabilistic
bisimulations for interesting examples.

\paragraph*{Expressiveness and Decidability.}
The name ``regular'' associated with this framework is because 
the set of formulas $\varphi(x_1,\ldots,x_k)$ first-order definable in $\univ$
coincides with \emph{regular relations}, i.e., relations definable by 
synchronous automata. More 
precisely, we define $\sem{\varphi}$ as the relation which contains all
tuples $(w_1,\ldots,w_k) \in (\ialphabet_\bot^*)^k$ such that
$\univ \models \varphi(w_1,\ldots,w_k)$. In addition,
we define the \defn{convolution} $w_1\otimes \cdots \otimes w_k$ 
of words 
$w_1,\ldots,w_k \in\ialphabet^*$ as a word $w$ over $\ialphabet_{\bot}^k$ (where
$\bot \notin \ialphabet$) such that $w[i] = (a_1,\ldots,a_k)$ with
\[
    a_j = \left\{ \begin{array}{cc}
    w_j[i] & \quad \text{if $|w_j| \geq i$, or} \\
    \bot   & \quad \text{otherwise.}
                \end{array}
        \right.
    \]
In other words, $w$ is obtained by juxtaposing $w_1, \ldots, w_k$ and padding
the shorter words with $\bot$. For example, $010 \otimes 00 =
(0,0)(1,0)(0,\bot)$. A $k$-ary relation $R$ over $\ialphabet^*$ is 
\defn{regular} if the set $\{ w_1 \otimes \cdots \otimes w_k : (w_1,\ldots,w_k)
\in R\}$ is a regular language over the alphabet $\ialphabet_{\bot}^k$. 
The relationship between $\univ$ and regular relations can be formally stated 
as follows.

\begin{proposition}[\cite{Bl99,BG04,BLSS03}]
\begin{enumerate}
\item
    Given a formula $\varphi(\bar x)$ over $\univ$, the relation
    $\sem{\varphi}$ is effectively regular. Conversely, given a regular
    relation $R$, we can compute a formula $\varphi(\bar x)$ over $\univ$
    such that $\sem{\varphi} = R$.
\item
    The first-order theory of $\univ$ is decidable.
\end{enumerate}
\end{proposition}

The decidability of the first-order theory of $\univ$ follows
using a standard automata-theoretic algorithm
(e.g. see \cite{Bl99,anthony-thesis}). 

In the sequel, we shall also use the term regular relations to denote relations
definable in $\univ$. In addition, to avoid notational clutter, we shall freely 
use other regular relations (e.g. successor relation $\prefSucc$ of the prefix 
$\pref$, and membership in a regular language) as syntactic sugar.

We note that the first-order theory of $\univ$ can also be reduced 
to weak monadic theory WS1S of one successor (therefore, allowing highly 
optimized tools like MONA 
\cite{monasecrets} and Gaston \cite{fiedor_lazy_2017})
by translating finite words to finite sets.
%
%
The relationship between the universal automatic
structure and WS1S can be made precise using the notion of \emph{finite-set 
interpretations} \cite{rubin-thesis,CL07}.

%% file: model.tex
\section{Probabilistic Bisimilarity within Regular Relations}
\label{sec:model}

In this section, we show how the framework of regular relations can be used
to encode a PTS, and the corresponding proof rules for probabilistic 
bisimulation.

\subsection{Specifying a probabilistic transition system}
Since we assume that all probability values specified in our systems
are rational numbers, the fact that our PTS is bounded-branching implies that 
we can
specify the probability values by natural \emph{weights} (by multiplying the
probability values by the least common multiple of the denominators). For
example, if a configuration $c$ has an action $toss$ that takes it to $c_1$ and
$c_2$, each with probability $1/2$, then the new system simply changes both 
values of $1/2$ to $1$. This is a known trick in the literature of 
probabilistic verification (e.g. see \cite{AHM07}). Therefore, we can now assume
that the transition probability functions have range $\Nat$.
\emph{The challenge now is that our encoding of a PTS in the universal automatic
structure must encode two different sorts} as words over a finite 
alphabet $\ialphabet$: configurations and natural weights.


Now we are ready to show how to specify a PTS $\struct$ in our framework.
Fix a finite alphabet $\ialphabet$ containing at least two letters 0 and 1.
We encode the domain of $\struct$ as words over $\ialphabet$.
In addition, a natural weight $n \in \Nat$ can be encoded in the usual way as a 
binary string. This motivates the following definition. 
\begin{definition}\label{def:regular-pts}
    Let $\struct$ be a PTS $\transysPTS$. We say that $\struct$ 
is \defn{regular} if the domain $S$ is a regular subset of $\ialphabet^*$
(i.e. definable by a first-order formula $\varphi(x)$ with one free variable 
over $\univ$), and if the graph of each function $\delta_a$ is a ternary regular
relation (i.e. definable by a first-order formula $\varphi(x,y,z)$ over
$\univ$, where $x$
and $y$ encode configurations, and $z$ encodes a natural weight).
\end{definition}

    Definition~\ref{def:regular-pts} is quite general since it allows for
    an infinite number of different natural weights in the PTS.
    Note that we can make do without the second sort (of numeric weights)
    if we have only finitely many numeric weights $n_1, \ldots, n_m$.
    This can be achieved by specifying a regular relation $R_{a,i}$
    for each action label $a \in \ACT$ and numeric weight $n_i$ with $i \in \NatInt{1}{m}$.

\begin{example}
\label{ex:random}
    We show a regular encoding of a very simple PTS: a random walk on the set of 
    natural numbers. At each position $x$, the system can non-deterministically
    choose to loop or to move. If the system chooses to loop, it will stay at the same
    position with probability 1. If the system chooses to move, it will
    move to $x+1$ with probability $1/4$, or
    move to $\max(0,x-1)$ with probability $3/4$.
    Normalising the probability values by multiplying by 4, we obtain the 
    numeric weights of 4, 1, and 3 for the aforementioned transitions,
    respectively. 
    
    To represent the system by regular relations, we encode
    the positions in unary and the numeric weights in binary.
    The set of configurations is the regular language $1^*$.
    The graph of the transition probability function can be described
    by a first-order formula
	$
        \varphi(x,y,z) := \varphi_{\rm loop}(x,y,z) \vee \varphi_{\rm move}(x,y,z)
    $
    over $\univ$, where
    \begin{eqnarray*}
        \varphi_{\rm loop}(x,y,z)   & := & x \in 1^* \wedge y \in 1^* \wedge ((x = y \wedge z = 100) \vee (x \neq y \wedge z=0)) \,;\\
        \varphi_{\rm move}(x,y,z) & := & x \in 1^* \wedge y \in 1^* \wedge ((x \prefSucc y \wedge  z = 1) \vee \\
            & & (y \prefSucc x  \wedge z = 11) \vee (x = \varepsilon \wedge y =\varepsilon \wedge z = 11) \vee \\
            & & (\neg(x \prefSucc y) \wedge \neg(y \prefSucc x)
                                \wedge \neg (x = \varepsilon \wedge y =\varepsilon) \wedge z = 0)). 
    \end{eqnarray*}\\[-5ex]
    \qed
\end{example}
\begin{example}
\label{ex:pPDA}
As a second example,
consider a PTS (from \cite{FJKW12}, Example 1)
described by a probabilistic pushdown automaton
with states $Q = \{p, q, r\}$ and stack symbols $\Gamma = \{X, X', Y, Z\}$.
There is a unique action $a$, and the transition rules $\delta_a$ are as follows:
\[
\begin{array}{llll}
pX \xrightarrow{0.5} qXX & \quad pX \xrightarrow{0.5} p &  \quad qX \xrightarrow{1} pXX & \quad rY \xrightarrow{1} rXX \\ 
rX \xrightarrow{0.3} rYX & \quad rX \xrightarrow{0.2} rYX' & \quad rX \xrightarrow{0.5} r \\
rX'\xrightarrow{0.4} r YX & \quad rX' \xrightarrow{0.1} rYX' & \quad rX' \xrightarrow{0.5} r
\end{array}
\]
A configuration of the PTS is a word in $Q \Gamma^*$,
consisting of a state in $Q$ and a word over the stack symbols. 
A transition can be applied if the prefix of the configuration matches the left hand side of the transition
rules above.
We encode the PTS as follows:
the set of configurations is $Q \Gamma^*$,
the weights are represented in binary after normalization, and
the transition relation $\varphi(x, y, z)$ encodes the transition rules in disjunction.
For example, the disjunct corresponding to the rule $\,pX \xrightarrow{0.5} qXX\,$ is
\begin{eqnarray*}
    x\in Q\Gamma^* \wedge y \in Q\Gamma^* \wedge (\exists u.~x = pXu \wedge y = qXXu) \wedge z = 101.
\end{eqnarray*}
Note that the PTS is bounded branching with a bound $3$.
\qed
\end{example}

%% file: model-words.tex
\subsection{Proof rules for probabilistic bisimulation}

Fix the set $\ACT$ of action symbols and the branching bound $N\ge 1$,
owing to the minimal deviation assumption.
Consider a two-sorted vocabulary $\sigma = \langle \{P_a\}_{a \in \ACT}, R, +\rangle$,
where $P_a$ is a ternary relation (with the first two arguments over the first sort, and
the third argument over the second sort of natural numbers), $R$ is a binary
relation over the first sort, and $+$ is the addition function over the
second sort of natural numbers.
The main result we shall show next is summarized in the following theorem:
\begin{theorem}
    There is a fixed first-order formula $\Phi$ over $\sigma$ such that
    a binary relation $R$ is a probabilistic bisimulation over a bounded-branching
    PTS $\struct = \transysPTS$ iff $(\struct,R) \models \Phi$. Furthermore,
    when $\struct$ is a regular PTS and $R$ is a regular relation,
    we can compute in polynomial time a first-order formula $\Phi'$
    over $\univ$ such that $R$ is a probabilistic bisimulation over $\struct$
    iff~$\univ \models \Phi'$.
    \label{th:verify}
\end{theorem}
This theorem implies the following result:
\begin{theorem}
    Given a regular relation $E \subseteq \ialphabet^* \times \ialphabet^*$ and a 
    bounded-branching regular PTS $\struct = \transysPTS$,
    there exists an algorithm that either finds $(u, v)\in E$ which are not probabilistically bisimilar 
    or finds a regular probabilistic bisimulation relation $R$ over $\struct$ such that $E \subseteq R$ if one exists.
    The algorithm does not terminate if{}f $E$ is contained in some 
    probabilistic bisimulation relation but
     every probabilistic bisimulation $R$ containing $E$ is not regular.
    \label{th:synthesis}
\end{theorem}

    Note that when verifying parameterized systems we are typically interested
    in checking bisimilarity over \emph{a set of pairs} (instead of just one 
    pair) of configurations, and hence $E$ in the above statement.

\begin{proof}[of Theorem \ref{th:synthesis}]
    To prove this, we provide two semi-algorithms, one for checking the
    existence of $R$ and the other for showing that a pair $(v,w) \in E$ is a
    witness for non-bisimilarity.

    By Theorem \ref{th:verify}, we can enumerate all possible candidate regular
    relation $R$ and effectively check that $R$ is a probabilistic
    bisimulation over $\struct$. The condition that $E \subseteq R$ is a
    first-order property, and so can be checked effectively.

    To show non-bisimilarity is recursively enumerable, observe that if we 
    fix $(v,w) \in E$ and a number $d$, then the restrictions $\struct_v$ and
    $\struct_w$ to configurations that are of distance at most $d$ away from
    $v$ and $w$ (respectively) are finite PTS. Therefore, 
    we can devise a semi-algorithm which enumerates all $(v,w) \in E$, and
    all probabilistic modal logic (PML) formulas \cite{LS91} $F$ over $\ACT$ 
    containing only
    rational numbers (i.e. a formula of the form $\langle a\rangle_\mu F'$,
    where $\mu \in [0,1]$ is a rational number, which is sufficient because
    we assume only rational numbers in the PTS). 
    We need to check that
    $\struct_v,v \models F$, but $\struct_w,w \nvDash F$. Model checking
    PML formulas over finite systems is decidable 
    (in fact, the logic is subsumed by Probabilistic CTL \cite{BaierKatoen}),
    which makes our check effective.
\end{proof}


\subsection{Proof of Theorem~\ref{th:verify}}\label{sec:verify}

In the rest of the section, we shall give a proof of Theorem \ref{th:verify}.
Given a binary relation $R \subseteq S \times S$, we can write a first-order 
formula $F_{\mathit{eq}}(R)$ for checking that $R$ is an equivalence relation:
\begin{align*}
\forall s,t,u\in S.\ & R(s,s) \wedge (R(s,t) \Rightarrow R(t,s)) \wedge ((R(s,t) \wedge R(t, u) \Rightarrow R(s,u)).
\end{align*}
We shall next 
define a formula $\varphi_a(p,q)$ for each $a \in \ACT$, such that $R$
is a probabilistic bisimulation for $\struct = \transysPTS$ iff
$(\struct,R) \models \Phi(R)$, where
\begin{equation}
\label{eq:phi}
    \Phi(R) := F_{\mathit{eq}}(R) \wedge \forall p,q \in S.~R(p,q) \Rightarrow \bigwedge\nolimits_{a \in \ACT}
    (\psi_a(p)\wedge \psi_a(q)) \vee \varphi_a(p,q).
\end{equation}
The formula $\psi_a(s) := \forall s' \in S.~\delta_a(s,s') = 0$ states that
configuration $s$ cannot move to any configuration through action $a$.

Before we describe $\varphi_a(p,q)$, we provide some intuition and define some intermediate
macros.
Fix configurations $p$ and $q$.
Informally, $\varphi_a(p,q)$ will first guess a set of configurations
$u_1,\ldots, u_N$ containing the successors of $p$ on action $a$,
and a set of configurations $v_1, \ldots, v_N$
containing the successors of $q$ on action $a$. 
Second, it will guess labellings $\alpha_1,\ldots,\alpha_N$ and $\beta_1,\ldots,\beta_N$
which correspond to partitionings of the configurations $u_1,\ldots, u_N$ and $v_1, \ldots, v_N$,
respectively.
The intuition is that the $\alpha$'s and $\beta$'s ``name'' the partitions:
if $\alpha_i = \alpha_j$ (resp.~$\beta_i = \beta_j$), then $u_i$ and $u_j$ (resp.~$v_i$ and $v_j$)
are guessed to be in the same partition.
The formula then checks that the guessed partitioning is compatible with the equivalence relation $R$
(i.e. if the labelling claims $u_i$ and $u_j$ are in the same partition, then indeed $R(u_i, u_j)$ holds),
and that the probability masses of the partitions assigned by configurations $p$ and $q$ 
satisfy the constraint given in~(\ref{defn:pb-proof-rule}).

For the first part, we define a formula
\begin{eqnarray*}
\mathsf{succ}_{a}(w; u_1,\ldots, u_N) & := & \left( \bigwedge\nolimits_{1\leq i < j\leq N} u_i \neq u_j \right) \wedge \\
 & & \quad\left( \forall u\in S.\ \delta_a(w, u) \neq 0 \Rightarrow \bigvee\nolimits_{1\leq i \leq N} u = u_i \right),
\end{eqnarray*}
stating that the successors of configuration $w$ on action $a$ are among the $N$
distinct configurations $u_1, \ldots, u_N$. 
Note that a configuration may have fewer than $N$ successors.
In this case, we can set the rest of the variables to arbitrary distinct configurations.

For the second part, we shall check that $R$ is compatible with the guessed partitions,
and that configurations $p$ and $q$ assign the same probability mass to the same partition.
Let $k_1,\ldots, k_n$ be a labelling for configurations $s_1,\ldots, s_n$.
To check that the partitioning induced by the labelling is compatible with $R$,
we need to express the condition that $k_i = k_j$ if and only if $R(s_i, s_j)$ holds.
To this end, we define a formula
\[
\mathsf{compat}_R(s_1,\ldots, s_n; k_1,\ldots,k_n) :=
\bigwedge\nolimits_{1\leq i<j \leq n} \left( R(s_i, s_j) \Leftrightarrow k_i = k_j \right).
\]
Now, we are ready to define $\varphi_a(p,q)$:
\begin{eqnarray}\label{formula:proof-rule}
\varphi_a(p,q)  & :=  &
  \exists u_1,\ldots, u_N, v_1,\ldots, v_N \in S.~\exists \alpha_1,\ldots,\alpha_N,
  \beta_1,\ldots,\beta_N \in\mathbb N. \nonumber\\
  & & \quad \mathsf{succ}_{a}(p; u_1,\ldots, u_N) \wedge \mathsf{succ}_{a}(q; v_1,\ldots, v_N) \wedge\\
  & & \quad \mathsf{compat}_R(u_1,\ldots, u_N,v_1,\ldots, v_N; \alpha_1,\ldots, \alpha_N,\beta_1,\ldots, \beta_N) \wedge \nonumber\\
  & & \quad \forall k \in \mathbb N.
    \left(\sum\nolimits_{i:~\alpha_i = k} \delta_a(p, u_i) = \sum\nolimits_{i:~\beta_i = k} \delta_a(q, v_i)\right). \nonumber
\end{eqnarray}
With this definition, $\varphi_a(p,q)$ holds if and only if     
$p \tran{\rho}_a S' \Leftrightarrow q \tran{\rho}_a S'$ holds
for any $\rho\ge 0$ and equivalence class $S' \in S/R$.
\OMIT{
\begin{proposition}\label{prop:prob-as-words-proof-rule}
    Let $\struct :=\transysPTS$ be a PTS
    where the branching is bounded by $N$.
    Given an equivalence $R$ over $S$,
    define formula $\varphi_a(p,q)$ as described above.
    Then $R$ is a probabilistic bisimulation for $\struct$ if and only if
    \begin{equation}
        \forall (p,q)\in R. \left(\bigwedge\nolimits_{a \in \ACT} \varphi_a(p,q) \right).
    \end{equation}
\end{proposition}
}

\begin{example}
Consider the PTS from Example~\ref{ex:pPDA}.
The configurations $pXZ$ and $rX$ are
probabilistic bisimilar.
This can be seen using a probabilistic bisimulation relation with equivalence classes 
$\{pX^kZ\} \cup \{rw : w \in \{X, X'\}^k\}$ for all $k\geq 0$
and 
$\{qX^{k+1}Z\} \cup \{rY w : w \in\{X, X'\}^k\}$ for all $k \geq 1$.
The probabilistic bisimulation relation is definable as the symmetric closure of
a regular relation $R$, where $(w_1, w_2)\in R$ iff
\begin{align*}
    & (w_1 = w_2)\ \vee\\
    & (w_1 \in pX^*Z \wedge w_2 \in r (X+X')^*\bot \wedge |w_1| = |w_2|)\ \vee\\
    & (w_1 \in r(X+X')^* \wedge w_2 \in r(X+X')^* \wedge |w_1| = |w_2|)\ \vee\\
    & (w_1 \in qX^*Z \wedge w_2 \in r Y(X+X')^*\bot \wedge |w_1| = |w_2|)\ \vee\\
    & (w_1 \in rY(X+X')^* \wedge w_2 \in rY(X+X')^* \wedge |w_1| = |w_2|).
\end{align*}
For this example, the formula~\eqref{eq:phi} simplifies to
$F_{\mathit{eq}}(R) \wedge \forall s,t\in S.~\varphi_a(p,q)$
for the unique action~$a$. 
This formula defines a condition that checks the bisimulation relation for all states symbolically.
To see the formula in action, fix configurations $pXZ$ and $rX$ which are probabilistic bisimilar.
In the PTS, $pXZ$ has two successors, $qXXZ$ and $pZ$, each with probability $0.5$,
and $rX$ has three successors, $rYX$ with probability $0.3$, $rYX'$ with probability $0.2$,
and $r$ with probability $0.5$.
In the formula for $\varphi_a(p,q)$, we can set the successors $u_i$ of $pXZ$ and 
the successors $v_j$ of $rX$ as above (the third ``successor'' $u_3$ is set 
to an arbitrary configuration not reachable from $pXZ$),
and set $\alpha_1 = 1$, $\alpha_2 = 2$, $\beta_1 = \beta_2 = 1$, and $\beta_3 = 2$,
corresponding to the equivalence classes of the bisimulation relation. 
One can check that the probability masses to these classes are the same.

We remark that the first-order theory of $\univ$ is sufficient to encode any probabilistic pushdown automaton, not
just this example.
\qed
\end{example}

We proceed to show that if $R$ and $\delta_a$
are first-order definable over $\univ$ then so are $\psi_a$ and $\varphi_a$. 
Suppose that $\delta_a$ is encoded using the ternary relation $\delta_a(x, y, z)$, as stated in the previous section.
(We shall re-use the symbol $\delta$ here to avoid a clash of names.)

We define $\psi_a(s) := \forall s'\in S.~\forall z\in \Nat.~\delta_a(s, s', z) \Leftrightarrow z=0$.
To define $\varphi_a$, the key is to express the sum of transition probabilities in the logic.
We use the fact that addition of integers in binary encoding is regular~(see e.g.~\cite{Bl99}),
and write a formula that performs iterated addition.
Formally, for each $a \in \ACT$ we define a formula $\chi_{a}$ such that
\begin{equation*}
\begin{split}
    & \chi_{a}(u; u_1,\ldots,u_N; \alpha_1,\ldots,\alpha_N; k; z) := \\
    & \qquad\exists z_1,\ldots,z_{N+1} \in \Nat.~z_1 = 0 \wedge z_{N+1} = z \wedge
    \bigwedge\nolimits_{1\le i \le N} \chi_{a}'(u, u_i, \alpha_i, k, z_i, z_{i+1}),
\end{split}
\end{equation*}
where
\begin{eqnarray*}
    \chi_{a}'(u, u', \kappa, k, x, y) & := &
    (\kappa = k \wedge \exists z.\ \delta_a(u, u', z) \wedge y = x + z) \vee (\kappa \neq k \wedge y = x)
\end{eqnarray*}
performs a single addition---we use the fact that addition ``$y = x+z$'' in binary is encodable
as a regular relation---and $z_1, \ldots, z_{N+1}$ store the intermediate sums.
Hence, given $k\in \mathbb N$, $u_1,\ldots,u_N,v_1,\ldots,v_N\in S$, and
$\alpha_1,\ldots,\alpha_N,\beta_1,\ldots,\beta_N\in \mathbb N$,
\begin{equation*}
    \sum\nolimits_{i:~\alpha_i = k} \delta_a(p, u_i) = \sum\nolimits_{i:~\beta_i = k} \delta_a(q, v_i)
\end{equation*}
if and only if
\begin{equation*}
    \exists z\in \Nat.\ \chi_{a}(p; u_1,\ldots,u_N; \alpha_1,\ldots,\alpha_N; k; z)
    \wedge \chi_{a}(q; v_1,\ldots,v_N; \beta_1,\ldots,\beta_N; k; z).
\end{equation*}
It follows that $\varphi_a(p,q)$ defined in (\ref{formula:proof-rule})
can be encoded in the first-order theory of $\univ$.
This concludes our proof of Theorem~\ref{th:verify}.
\OMIT{
\begin{eqnarray*}
\varphi_a(p,q)  & :=  &
  \exists u_1,\ldots, u_N\in S.\ \exists v_1,\ldots, v_N \in S.\\
  & & \exists \alpha_1,\ldots,\alpha_N \in\mathbb N.\
    \exists \beta_1,\ldots,\beta_N \in\mathbb N. \\
    & & \quad \mathsf{succ}_{a}(p; u_1,\ldots, u_N) \wedge \mathsf{succ}_{a}(q; v_1,\ldots, v_N) \wedge \\
    & & \quad \mathsf{compat}_R(u_1,\ldots, u_N,v_1,\ldots, v_N; \alpha_1,\ldots, \alpha_N,\beta_1,\ldots, \beta_N) \wedge\\
    & & \quad \bigwedge\nolimits_{1\leq k \leq 2N}
    \left(\exists z\in \Nat.\ 
     \begin{array}{l}
        \chi_{a}(p; u_1,\ldots,u_N; \alpha_1,\ldots,\alpha_N; z, k)\ \wedge \\
        \chi_{a}(q; v_1,\ldots,v_N; \beta_1,\ldots,\beta_N; z, k)
     \end{array}
\right).
\end{eqnarray*}
}

\paragraph{Remark.}
Note that it is decidable to check whether a given presentation of a regular PTS is valid.
To see this, suppose that a set $\Delta := \{\delta_a(x,y,z)\}_{a\in\ACT}$
of formulae is claimed to encode 
the probabilistic transition functions of a PTS with a branching bound $N$.
Fix a formula $\delta_a \in \Delta$. First, we need to check that
for all $x\in S$, there are at most $N$ distinct $y$'s
such that $\delta_a(x,y,z)$ satisfies $z \neq 0$. 
Second, we need to check that $\sem{\delta_a}$
is a function, i.e., $\forall x,y.~\exists !z.~\delta_a(x,y,z)$, where $\exists
!z.~\varphi(\bar x,z)$ is a shorthand for the formula asserting there exists
precisely one $z$ such that $\varphi(\bar x,z)$ is true.
Third, we need to check that $\sem{\delta_a}$ encodes a mapping
$S \to \{\zerodist \} \cup \mathcal{D}_S$.
The first two requirements are easily seen to be expressible as a first-order
formula and hence is algorithmic over $\univ$.
The third requirement amounts to checking the assertion
that there exists $w_a\in \mathbb N$ satisfying 
\begin{eqnarray*}
    & & ~\forall x\in S.~
         (\forall y\in S.~\forall z\in \Nat.~\delta_a(x, y, z) \Leftrightarrow z=0) \vee \\
    & & \qquad \qquad ( \exists y_1,\dots,y_N\in S.~\exists z_1,\dots,z_N\in \mathbb N.\\
    & & \qquad \qquad \qquad \mathsf{succ}_{a}(x; y_1,\ldots, y_N) \wedge
          \bigwedge\nolimits_{1\le i \le N} \delta_a(x, y_i, z_i) \wedge
          \sum\nolimits_{1\le i \le N} z_i = w_a),
\end{eqnarray*}
which is a first-order formula and is algorithmic over $\univ$
by the fact that summation of a fixed number of weights is regular
(as shown earlier in this section).
Finally, since all of the $w_a$'s are expected to 
be the same common multiple of the denominators
of the transition probabilities, we need to check that
there is $w\in \mathbb N$ such that $w_a = w$ for all $a \in \ACT$.
This is again algorithmic as we can pinpoint the exact value of each $w_a$ by enumeration.

%% file: case-study.tex
\section{Application to Anonymity Verification}
\label{sec:application}

In this section, we show how to verify the anonymity property of
cryptographic protocols via computation of probabilistic bisimulations.
We shall first formalize the connection between
the concepts of anonymity and probabilistic bisimulation.
We then introduce a verification framework and apply it to verify
the anonymity property of the dining cryptographers protocol~\cite{chaum1988dining}
and the grades protocol~\cite{apex}.

Fix a PTS $\struct := \transysPTS$.
Let $\pi(\struct)$ denote the set of all finite paths of~$\struct$.
An \emph{adversary} $f: \pi(\struct) \to \mathcal{D}_{\ACT}$
resolves the nondeterministic choices of $\struct$ and
induces a PTS
$\struct_f := \langle S', \{\delta'_a\}_{a \in \ACT} \rangle$
where $S' := \pi(\struct) \uplus \{\bot \}$,
and $\delta'_a : S' \to \mathcal{D}_{\ACT}$
is defined such that for each finite path
$\pi := s_0 \to_{a_1} \cdots \to_{a_n} s_n$
and
$\pi' := s_0 \to_{a_1} \cdots \to_{a_n} s_n \to_{a} s_{n+1}$ in $\pi(\struct)$,
\vspace{-.5em}
\[
    \delta'_a(\pi, \pi') := f(\pi)(a) \cdot \delta_a(s_n, s_{n+1}).
\]
Furthermore,
$\delta'_a(\bot, \bot) := 1$ and
$\delta'_a(\pi, \bot) := 1 - \sum_{\pi' \in \pi(\struct)} \delta'_a(\pi, \pi')$
for all $\pi \in \pi(\struct)$.
Intuitively,
$\struct_f$ describes the behaviour of $\struct$
under the adversary $f$
by successively extending the paths of $\struct$ according to the distributions selected by $f$.
The special ``sink state`` $\bot$ indicates
that $f$ has got stuck and thus the path can no longer be extended.
The paths of $\struct_f$ therefore induce a probability measure
that can be formalised using the standard cylinder construction,
see e.g.~\cite{marta-survey}.
More precisely,
a finite path $\ModelRun := \pi_0 \to_{a_1} \cdots \to_{a_n} \pi_n$ of $\struct_f$
defines a \emph{basic cylinder} $Run_{\ModelRun}$,
which is the set of all finite/infinite paths with $\ModelRun$ as a prefix.
We associate this cylinder with probability
$
\Pr^{\pi_0}(Run_{\ModelRun}) := \prod_{i=1}^{n} \delta_{a_i}'(\pi_{i-1}, \pi_{i}).
$
This definition gives rise to a unique probability measure for the $\sigma$-algebra
over the set of all paths from $\pi_0$.
We define $\tau(\ModelRun) := a_1 \cdots a_n$ to be the \emph{trace} of $\ModelRun$,
and call a set $\mathcal T \subseteq \ACT^+$ a \emph{trace event}.
The probability of a trace event $\mathcal T$ with respect to a configuration $s \in S$ is given by
\[
\Pr\nolimits^{s}(\mathcal T) :=
\Pr\nolimits^{s}(\bigcup \left\{Run_\pi : \tau(\pi)\in\mathcal T, \mbox{$\ModelRun$ starts from $s$} \right\}).
\]

Now we are ready to define the concept of anonymity.
Let $\mathcal I \subseteq \transysDom$
be a set of initial configurations.
We say that $\struct$ is \emph{anonymous to an adversary $f$}
if for all $s \in \mathcal I$ and trace event $\mathcal T$,
the value of $\Pr^s(\mathcal T)$ is solely
determined by $\mathcal T$, and thus independent on $s$.
Intuitively, this means that
the adversary cannot obtain any information about a specific
initial configuration by experimenting on the system and
observing the traces.

We shall only consider external adversaries in our case study.
An adversary $f: \pi(\struct) \to \mathcal{D}_{\ACT}$ is \emph{external}
if $f(s) = f(s')$ for all $s,s' \in S$, and
$f(s_0 \to_{a_1} \cdots \to_{a_n} s_n) = f(s_0' \to_{a_1'} \cdots \to_{a_n'} s_n')$
when $a_i = a_i'$ for $i \in \NatInt{1}{n}$.
That is, an external adversary
resolves a nondeterministic choice solely
based on the trace she has observed so far.
We call a PTS \emph{anonymous} if it is anonymous to any external adversary.
The following result establishes a connection between
the anonymity property and probabilistic bisimulations.

\begin{proposition}\label{prop:anonymity}
    Let $\struct := \transysPTS$ be a PTS and
    $f$ be an external adversary for $\struct$.
    Then for all $u,v \in \transysDom$ such that $u \sim v$,
    $\Pr^u(\mathcal T) = \Pr^v(\mathcal T)$ holds for
    any trace event $\mathcal T$ in $\struct_f$.
    That is, configurations $u$ and $v$ induce the same trace distribution in $\struct_f$.
\end{proposition}

Based on Proposition~\ref{prop:anonymity},
we propose a framework to verify the anonymity property of $\struct$ as follows.
We first specify a ``reference system''
$\struct' := \langle S; \{\delta_a' \}_{a\in\ACT}\rangle$
that has the same initial configurations and actions as those of $\struct$,
except that the trace distribution of $\struct'_f$ is independent of
specific initial configurations for any adversary $f$.
We then explore bisimulation relations $R$ for $\struct \sqcup \struct'$
satisfying $R \supseteq \{ (s,s') \in \mathcal I \times \mathcal I'  :  s = s' \}$.
When such a relation $R$ is found, we can conclude 
that the trace distribution of $\struct_f$ is also independent of the initial configurations
for any adversary $f$, thereby proving the anonymity property of $\struct$.

\subsubsection{The dining cryptographers protocol.}
Dining cryptographers protocol~\cite{chaum1988dining}
is a multi-party computation algorithm aiming to securely
compute the XOR of the secret bits held by the participants.
More precisely, consider a ring of $n\ge 3$ participants $p_0,\dots,p_{n-1}$
such that each participant $p_i$ holds a secret bit $x_i$.
To compute $x_0 \oplus \cdots \oplus x_{n-1}$
without revealing information about the values of $x_0,\ldots,x_{n-1}$,
the participants carry out a two-stage computation as follows:
i) Each two adjacent participants $p_i$, $p_{i+1}$ compute a random bit $b_i$ that is accessible only to them;
ii) Each participant $p_i$ announces the value $a_i := x_i \oplus b_i \oplus b_{i-1}$\footnote{
All arithmetical operations on the subscripts are performed modulo $n$
to take the ring structure into account.
} to the other participants.
Hence, every participant $p_i$ can observe the values of $x_i$, $b_i$,  $b_{i-1}$ and $a_0, \ldots, a_{n-1}$.
It turns out that
$a_0 \oplus \cdots \oplus a_{n-1} = x_0 \oplus \cdots \oplus x_{n-1}$,
so all participants are able to compute the XOR of the secret bits
after executing the protocol.
Furthermore,
the anonymity property of the protocol assures that
any individual participant $p_i$ cannot infer the values of
the other secret bits from the information she has observed during the execution of the protocol.

We model the protocol as a length-preserving regular PTS.
The configurations of a ring of $n$ participants are encoded as words of size $n$.
The initial configurations are words $w \in \{0,1\}^*$ such that $w[i]$ represents $x_i$
for $i \in \NatInt{0}{|w|-1}$.
The transition relation consists of six transitions:
observer non-deterministically tossing head (via action $\head$),
observer non-deterministically tossing tail (via action $\tail$),
non-observer tossing head with probability 0.5 (via action $\toss$),
non-observer tossing tail with probability 0.5 (via action $\toss$),
participant announcing zero (via action $\zero$), and
participant announcing one (via action $\one$).
The outcomes of the tosses by the observer are visible
(i.e.~as actions $\head$ and $\tail$), while
the outcomes of the tosses by the other participants are hidden
(i.e.~as action $\toss$).
Each maximal trace from an initial configuration of size $n$
consists of $n$ successive tossing actions,
followed by $n$ successive announcing actions.
Starting from an initial configuration $w$ and for $i \in \NatInt{0}{n-1}$,
the $i$-th toss action updates the value of
$w[j]$ to $w[j] \oplus b_i$ for $j \in \{i, {i+1}\}$,
where $b_i = 1$ if a head is tossed and $b_i=0$ otherwise.
Any configuration $v$ reached after $n$ tosses
would satisfy $v[i] = x_i \oplus b_i \oplus b_{i-1}$
for $i \in \NatInt{0}{n-1}$.
The PTS then ``prints out'' the configuration
by going through $n$ announcement transitions via actions
$a_0, \dots, a_{n-1}$,
such that $a_i$ is $\one$ if $v[i] = 1$ and $a_i$ is $\zero$ if $v[i] = 0$.

We consider the case where the first participant of the protocol
is the observer.
The maximal traces of the PTS in this case are
in form of $t \cdot t'$, where $|t| = |t'|$, 
$t \in \{\head, \tail\}\,\toss^*\{\head, \tail\}$,
and $t' \in \{\zero,\one\}^*$.
For example, 
$\head$ $\toss$ $\tail$ $\one$ $\zero$ $\zero$
is a maximal trace starting from initial configuration $010$.
To prove anonymity, we define a reference system
such that the initial configurations and the actions are
the same as those of the original PTS,
except that the announcements $a_0, \dots, a_{n-1}$
encoded in the maximal traces from an initial configuration $w$
are uniformly distributed over 
$\{ (a_0, \dots, a_{n-1}) : a_0 \oplus \cdots \oplus a_{n-1} = w[0] \oplus \cdots \oplus w[n-1],
~a_0 = w[0] \oplus b_0 \oplus b_{n-1}\}$.\footnote{
Such a distribution can be obtained
by i) choose $a_1, \dots, a_{n-2}\in \{0,1\}$ uniformly at random;
ii) set $a_0 = w[0] \oplus b_0 \oplus b_{n-1}$;
iii) set $a_{n-1} = a_0 \oplus \cdots \oplus a_{n-2} \oplus w[0] \oplus \cdots \oplus w[n-1]$.}
In this way, the distribution of the announcements is independent of
the initial configuration once the values of
$x_0 \oplus \cdots \oplus x_{n-1}$, $x_0$, $b_0$, and $b_{n-1}$
(i.e. the information revealed to the first participant) are fixed.
We then compute a probabilistic bisimulation between
the original system and the reference system,
establishing the anonymity property that the first participant
cannot infer the secret bits of the other participants
from the information she observes.

\paragraph{A generalized dining cryptographers protocol.}
We have also considered a generalized dining cryptographers protocol
where the secrets $x_0, \ldots, x_{n-1}$ of the $n$ participants
are bit-vectors of the same size.
Note that the set of the initial configurations is not regular
when the size of the secrets is parameterized.
To construct a regular model, we allow a configuration to encode
secrets of different sizes, and
devise the transition system such that
an initial configuration $w$ can finish the protocol
(i.e.~can have a trace containing all of the announcements $a_0, \ldots, a_{n-1}$)
if and only if the messages encoded in $w$ have same size.
The resulting PTS is a regular system;
it over-approximates the PTS of the generalized
dining cryptographers protocol in the sense that
the anonymity property of the former implies that of the latter.
More details of the model can be found in Appendix~\ref{sec:gDCP}.

\subsubsection{The grades protocol.}

The grades protocol~\cite{apex}
is a multi-party computation algorithm aiming to securely
compute the sum of the secrets held by the participants.
The setting of the protocol is pretty similar to that of the
dining cryptographers:
given $n\ge 3$ and $g \ge 2$,
we have a ring of $n$ participants $p_0, \ldots, p_{n-1}$
where each participant $p_i$ holds a secret $x_i \in \NatInt{0}{g-1}$.
Note that both $g$ and $n$ are parameterized in this protocol.
The goal of the participants is to compute the sum $x_0 + \cdots + x_{n-1}$
without revealing information about the individual secrets.
Define $M := (g-1) \cdot n + 1$.
The protocol consists of two steps:
i) Each two adjacent participants $p_i$, $p_{i+1}$
compute a random number $y_i \in \NatInt{0}{M-1}$;
ii) Each participant $p_i$ announces
$a_i := (x_i + y_i - y_{i-1})~{\rm mod}~M$
to the other participants.
After executing the protocol, the participants compute $a := a_0 + \cdots + a_{n-1}~{\rm mod}~M$.
Because of the ring structure, the $y_i$'s will be cancelled out in the sum.
Thus the value of $a$ will equal to the sum of all secrets.
The anonymity property of the protocol asserts that no participant can infer
the secrets held by the other participants from the information she has observed.

We consider a variant of the grades protocol where $M$ can be
any power of two greater than $(g-1) \cdot n$.
Observe that the same anonymity and correctness property
of the original protocol also holds for this variant.
To verify the anonymity property, we model an over-approximation of the protocol
where the secrets are allowed to range over $\NatInt{0}{M-1}$.
This model
is similar to the one we have constructed for the generalized dining cryptographers protocol
except that, e.g., the XOR operations are now replaced with bitwise additions and negations.
A reference system is specified such that
the announcements $a_1, \ldots, a_{n-1}$ observed by the first participant $p_0$
are uniformly distributed over the values 
satisfying $a_0 + \cdots + a_{n-1}~{\rm mod}~M
= x_0 + \cdots + x_{n-1}~{\rm mod}~M$.
By computing a probabilistic bisimulation between 
the original system and the reference system,
we establish the anonymity property
that the grades protocol is anonymous whenever $M$
is chosen as a power of two with $(g-1) \cdot n < M$.
See Appendix~\ref{sec:grades} for model details of the model. 

%% file: learning.tex
\section{Learning Probabilistic Bisimulations}
\label{sec:learning}

We propose an automata learning method to automatically compute
regular probabilistic bisimulations~$R$, focusing on the case of
\emph{length-preserving} PTSs, which covers all examples given in
the previous section. The approach uses active automata
learning, for instance Angluin's $L^*$
method~\cite{angluin1987learning} or refinements of it, to compute
$R$. This approach is inspired by previous work on using active automata learning
for invariant inference~\cite{DBLP:conf/icfem/VardhanSVA04,CHLR17}.
Our procedure assumes
\begin{inparaenum}[(i)]
\item as input a bounded-branching PTS~$\struct = \transysPTS$, as
  well as a length-preserving regular
  relation~$E \subseteq (\ialphabet \times \ialphabet)^*$ supposed to
  be covered by $R$;
\item an effective way to check the correctness of $R$, i.e., a
  decision procedure in the sense of Theorem~\ref{th:verify}; and
\item a procedure to compute the greatest probabilistic
  bisimulation~$\bar R_n \subseteq (\ialphabet \times \ialphabet)^n$
  for $\struct$ restricted to configurations of any length~$n \in \mathbb N$.
\end{inparaenum}
The last assumption can easily be satisfied for length-preserving
PTSs. Indeed, such systems, restricted to configurations of length~$n$,
are finite-state, so that efficient existing
methods~\cite{Baier96,DHS03,VF10,CBW12} apply.
A solution $R$ is presented as a deterministic letter-to-letter transducer,
i.e., as a deterministic finite-state automaton over the alphabet
$\ialphabet \times \ialphabet$. 

\begin{algorithm}[tb]
    \KwIn{Candidate automaton~$\cal H$ over $\ialphabet \times \ialphabet$, PTS~$\struct$,
        and regular relation $E \subseteq (\ialphabet \times \ialphabet)^*$.}
    \KwResult{%
        \makebox[9em][l]{$\mathit{NoSolution}(v, w)$}
        if there is no bisimulation~$R$ with $E \subseteq R$;
        \makebox[9em][l]{$\mathit{PositiveCEX}(v, w)$}
        if $\cal H$ should accept $(v, w)$, but
        does not;\newline
        \makebox[9em][l]{$\mathit{NegativeCEX}(v, w)$}
        if $\cal H$ accepts $(v, w)$, but
        should not;\newline
        \makebox[9em][l]{$\mathit{Correct}$}
        if $\cal H$ encodes a correct
        bisimulation for~$\struct$ and $E \subseteq {\cal L}({\cal H})$.
    }
    \medskip   
        \smallskip
        Check whether $E \subseteq {\cal L}({\cal H})$, and
        whether $\struct \models \Phi({\cal L}({\cal H}))$ using the $\Phi$
        from \eqref{eq:phi}\;
        \uIf{there is a counterexample of minimal length $n$}{
            Compute the greatest bisimulation
            $\bar R_n$ restricted to configurations of length $n$\;
              \uIf{there is $(v , w) \in E \setminus \bar R_n$ with $|v| = |w| = n$}{
                   Output $\mathit{NoSolution}(v ,w)$ and abort\;
              }
              \ElseIf{there is $(v , w) \in {\cal L}({\cal H}) \setminus \bar R_n$ with $|v| = |w| = n$}{
                   \Return{$\mathit{NegativeCEX}(v, w)$}\;
              }
              \ElseIf{there is $(v , w) \in \bar R_n \setminus {\cal L}({\cal H})$}{
                   \Return{$\mathit{PositiveCEX}(v, w)$}\;
          }        
        }
        \Else{
             \Return{$\mathit{Correct}$}\;
        }

    \caption{Equivalence check for $L^*$}
    \label{alg:eqv}
\end{algorithm}

Since $L^*$-style learning requires the taught language to be uniquely
defined, our approach attempts to learn a representation of the
greatest \emph{length-preserving} probabilistic bisimulation relation
$\bar R \subseteq (\ialphabet \times \ialphabet)^*$, which is the
unique bisimulation relation formed by the union of all
length-preserving probabilistic bisimulations of~$\struct$, i.e.,
$\bar R = \bigcup_{n\ge 1} \bar R_n$.  Because $\bar R$ is not in
general computable, the learning process might diverge and fail to
produce any probabilistic bisimulation. It can also happen
that learning terminates, but yields a probabilistic bisimulation
relation strictly smaller than $\bar R$.
\OMIT{Our approach is guaranteed to
terminate whenever the greatest bisimulation $\bar R$ is regular;
the produced relation~$R$ is a probabilistic bisimulation relation
(since it satisfies the proof rule based on Theorem~\ref{th:verify})
that might be strictly smaller than $\bar R$.}

The $L^*$ method requires a teacher that is able to answer two kinds of queries:
\begin{itemize}
\item \textbf{membership queries}, i.e., whether a
  pair~$(v, w)$ of words should
  be accepted by the automaton to be learned. Since our learner tries
  to learn the greatest bisimulation, the teacher can answer this query
  by checking whether the configurations~$v, w$ are bisimilar; this is
  done by computing the greatest bisimulation~$\bar R_{|v|}$
  restricted to configurations of any length~$|v| = |w|$, and checking
  whether or not $(v, w) \in \bar R_{|v|}$.
\item \textbf{equivalence queries}, i.e., whether a candidate
  automaton~$\cal H$ is the correct language to be learned. Such
  queries can essentially be answered by checking whether the language
  ${\cal L}({\cal H})$ satisfies the formula~$\Phi(R)$ from \eqref{eq:phi}.
  The complete algorithm for answering equivalence queries
  is given in Algorithm~\ref{alg:eqv}. 
  The algorithm first attempts to find a shortest counterexample
  to the proof rule.
  If a counterexample of length~$n$ is found, then
  the difference set ${\cal L}({\cal H}) \,\Delta\, \bar R_n$ must contain
  at least one pair of length~$n$.
  Any of such pairs is a valid counterexample for automata learning
  since the learner tries to learn the greatest bisimulation.
  The teacher thus reports one such pair to be a positive or negative counterexample
  according to its membership in $\bar R_n$.
\end{itemize}

\paragraph{Properties of the Learning Algorithm.}
The learning procedure terminates
when the teacher outputs \emph{NoSolution} or returns \emph{Correct}
for an equivalence query.
In the former case, the teacher explicitly provides a pair of non-bisimilar configurations in $E$.
In the latter case, the procedure computes an automaton $\cal H$
such that $E \subseteq {\cal L}({\cal H})$ and ${\cal L}({\cal H})$
is a correct probabilistic bisimulation
(as it satisfies the proof rule based on Theorem~\ref{th:verify}), though
not necessarily the greatest one.
Since all counterexamples reported by the teacher
are contained in ${\cal L}({\cal H}) \,\Delta\, \bar R$,
the learning procedure is guaranteed to terminate
for PTSs where the greatest probabilistic bisimulation $\bar R$ is regular.

\OMIT{
\paragraph{Restriction to Reachable Configurations.}
For PTS~$\struct = \transysPTS$ with a \emph{regular} set~$C$ of
reachable configurations, and assuming $E \subseteq C \times C$, there
is a natural way to optimize the learning procedure by only
considering the reachable configurations. This is done by simply
replacing the greatest finite-length bisimulations~$\bar R_i$ in
Algorithm~\ref{alg:eqv}, and when answering membership queries, with the
greatest bisimulation $\bar R_i^c = \bar R_i \cap C$ on the reachable
configurations; since $\bar R_i^c$ can be a lot smaller than $\bar
R_i$, this can lead to significant speed-ups. Note that a
bisimulation~$R'$ on reachable configuration can be extended to a
bisimulation~$R$ on all configurations by setting~$R = R' \cup \{(v,
v) : v \not\in C\}$.
An over-approximation of the set~$C$ can be computed using techniques like
in \cite{DBLP:conf/icfem/VardhanSVA04,CHLR17}.

\chihduo{How about replacing the above paragraph with the following one?
The set of reachable configurations is not regular for the generalized DCP
and the grades protocol, but in our experiments, we do reduce the size of the
candidate bisimulations with a user-specified inductive invariant called {\sf isConfig}.}
}
\paragraph{Optimization with Inductive Invariants.}
There is a natural way to
optimize the learning procedure by only considering
a \emph{regular} inductive invariant $\mathit{Inv}$
such that $\mathit{Inv}$ contains the set of reachable configurations
and $E \subseteq \mathit{Inv}\times \mathit{Inv}$. The optimization is done by simply
replacing the greatest finite-length bisimulations~$\bar R_i$ in
Algorithm~\ref{alg:eqv}, and when answering membership queries, with the
greatest bisimulation $\bar R_i^I = \bar R_i \cap \mathit{Inv}$ on the
inductive invariant. Since $\bar R_i^I$ can be a lot smaller than $\bar
R_i$, this can lead to significant speed-ups. Note that a
bisimulation~$R'$ on $\mathit{Inv}$ can be extended to a
bisimulation~$R$ on all configurations by setting~$R = R' \cup \{(v,
v) : v \not\in \mathit{Inv}\}$. The inductive invariant~$\mathit{Inv}$ may be
manually specified, or automatically generated using techniques like
in \cite{DBLP:conf/icfem/VardhanSVA04,CHLR17}.

\paragraph{Experimental Results and Conclusion.}
We have implemented a prototype\footnote{
Available at https://bitbucket.org/chihduo/prob-bisim-tool/src/master/.}
in Scala to evaluate our learning method.
Given a PTS specified over $\univ$, we manually translate it to WS1S formulas
(see~Appendix \ref{sec:univ-to-ws1s})
and obtains finite automata for these formulas using the Mona tool~\cite{klarlund2001mona}.
Our prototype then applies the $L^*$ learning procedure as described in this section,
including the optimization to consider only the configurations of valid format.
When answering an equivalence query,
our tool invokes Mona to verify candidate automata
and obtain counterexamples (line~1-2 of Algorithm~\ref{alg:eqv}).
We have used the prototype tool to check the anonymity property of the
three protocols described in Section~\ref{sec:application}.
The proofs generated by our tool 
are finite automata encoding the witness probabilistic bisimulation relations.
We summarize the experimental in Table~\ref{tab:experiment}.
\input{table}

%% file: table.tex
\begin{table}
\centering
\begin{tabular}{|c||c|c||c|c|c|}
\hline 
Case study & \#states & \#trans & mona & bisim & total\tabularnewline
\hline 
\hline 
Dining Cryptographers, single-bit & 13 & 832 & 2s & 2s & 6s \tabularnewline
\hline 
Dining Cryptographers, multi-bit & 16 & 1024 & 3s & 24s  & 28s \tabularnewline
\hline 
The grades protocol & 25 & 1600 &  5s & 28s & 35s \tabularnewline
\hline 
\end{tabular}
\vspace{1em}
\caption{Experimental results. For each case study,
    we list the size of the final proof produced by our tool, 
    the time taken by Mona to verify the candidate automata,
    the time taken by our tool to compute the fixed-length bisimulations,
    and the total computation time of the learning procedure.
    Experiments are run on a Windows laptop with 2.4GHz Intel i5 processor
    and 2GB memory limit. }
\label{tab:experiment}
\end{table}

%% file: appendix.tex
\section{Appendix}

\subsection{Translating $\operatorname{FO}(\mathfrak U)$ to WS1S}
\label{sec:univ-to-ws1s}
In this section, we show how to translate the first-order theory of the universal structure
$\univ := \univDesc$ to the weak monadic second-order logic of one successor (WS1S).
It suffices to consider $\Sigma = \{0,1\}$,
since any alphabet $\Sigma$ can be encoded in binary
on blocks of uniform length $\lceil{\lg |\Sigma|}\rceil$.
We define an interpretation of $\univ$ in WS1S as
$
I(\mathfrak U) := \langle \Nat, {Pref}, {EqL}, \{L_a\}_{a\in\Sigma}\rangle,
$
where
\begin{eqnarray*}
    {Pref}(X,Y) & := & 
    \exists x.\exists y.\:m(x,X) \wedge m(y,Y)
    \wedge x \le y \wedge \forall z<x.\:z\in X \Leftrightarrow z,\\
    {EqL}(X,Y) & := & \exists x.\exists y.\:m(x,X)\wedge m(y,Y)\wedge x = y,\\
    L_1(X)  & := & \exists x.\:m(x, X) \wedge x\ge 1 \wedge x-1,\\
    L_0(X)  & := & \exists x.\:m(x, X) \wedge x\ge 1 \wedge x-1 \notin X,
\end{eqnarray*}
and $m(x,X) := x\in X\wedge \forall x'\in X.~x'\le x$
is a macro meaning that $x$ is the maximal element of $X \subseteq \Nat$.
Now, consider an isomorphic mapping $f: \Sigma^*\to \mathcal{F}(\Nat)$,
where $\mathcal{F}(\Nat)$ denotes the set of the finite subsets of $\Nat$,
by $f(w) := \{n\in\Nat : (w \cdot 1)[n+1] = 1\}$.
That is, $f$ maps $w$ to a set
whose characteristic string is $w \cdot 1$ (recall that word indices start from one).
It is straightforward to show that for $w,w'\in\Sigma^*$ and $a \in \Sigma$,
it holds that $w \preceq w'$ iff ${Pref}(f(w),f(w'))$,
$\eqlen(w,w')$ iff ${EqL}(f(w),f(w'))$, and
$l_a(w)$ iff $L_a(f(w))$.
Therefore, $I(\mathfrak U)$ is indeed an interpretation of $\univ$.

\subsection{Modelling the generalized dining cryptographers protocol}
\label{sec:gDCP}
Given parameters $n\ge 3$ and $m\ge 1$,
consider a ring of $n$ participants $p_0, \ldots, p_{n-1}$ where 
each participant $p_i$ possesses a secret bit-vector $x_i$ of size $m$.
The protocol consists of two stages.
At stage one, each pair of neighboring participants $p_i$ and $p_{i+1}$\footnote{
Note that all arithmetic operations on the subscripts are performed modulo
the number of the participants in the protocol.}
computes a random vector $b_i$ by tossing a coin for $m$ times.
At stage two, each participant $p_i$ announces a vector
$a_i := x_i \oplus b_i \oplus b_{i-1}$
to the other participants.
After the protocol is carried out,
the participants compute the XOR of the secrets
as $a_0 \oplus \cdots \oplus a_{n-1} = x_0 \oplus \cdots \oplus x_{n-1}$.
Note that both $n$ and $m$ are parameterized in this protocol.

We model the generalized dining cryptographers protocol as a regular PTS.
The set of the initial configurations is
$\,\mathcal I := \{0,1\}^* (\texttt{\#} \{0,1\}^*)^*$.
Given an initial configuration
$w := w_0\texttt{\#}\dots\texttt{\#}w_{m-1} \in \mathcal I$,
define $N(w) := \min_{0 \le i < m} |w_i|$.
The configuration $w$ encodes secret bit-vectors
$x_0, \ldots, x_{N(w)-1}$ of size $m$ for a ring of $N(w)$ participants.
The encoding is specified such that $w_k[i] = x_i[k]$
for $k \in \NatInt{0}{m-1}$ and $i \in \NatInt{0}{N(w)-1}$.
In other words, the first $N(w)$ bits of segment $w_k$
are $x_0[k], \ldots, x_{N(w)-1}[k]$.

The transition relation consists of seven transitions:
observer non-deterministically tossing head (via action $\head$),
observer non-deterministically tossing tail (via action $\tail$),
non-observer tossing head with probability 0.5 (via action $\toss$),
non-observer tossing tail with probability 0.5 (via action $\toss$),
participant finishing computing a random vector (via action $\reset$),
participant announcing zero (via action $\zero$), and
participant announcing one (via action $\one$).
The outcomes of the tosses by the observer are visible
(i.e.~as actions $\head$ and $\tail$), while
the outcomes of the tosses by the other participants are hidden
(i.e.~as action $\toss$). Recall that a participant
needs $m$ successive tosses to compute a random vector of size $m$.
We would refer to such a succession of tosses as a \emph{round}
and stipulate that an $\reset$ action must be made after each round.

Fix an arbitrary initial configuration $w := w_0\texttt{\#}\dots\texttt{\#}w_{m-1}$
and let $n := N(w)$.
A maximal trace from configuration $w$
begins with $n$ rounds (each followed by an $\reset$ action)
and ends with $n m$ successive announcements.
For each $i \in \NatInt{0}{n-1}$ and $k \in \NatInt{0}{m-1}$,
the $k$-th toss made in the $i$-th round
updates $w_k[j]$ to $w_k[j] \oplus b_i^k$ for $j \in \{i, i+1\}$,
where $b_i^k := 1$ if a head is tossed and $b_i^k := 0$ otherwise.
The configuration $u := u_0\texttt{\#}\dots\texttt{\#}u_{m-1}$
reached from $w$ after $n$ rounds of tosses will satisfy
$u_k[i] = x_i[k] \oplus b_i^k \oplus b_{i-1}^k$
for $i \in \NatInt{0}{n-1}$ and $k \in \NatInt{0}{m-1}$.
Hence for each $i \in \NatInt{0}{n-1}$,
\begin{eqnarray*}
(u_0[i], \ldots, u_{m-1}[i]) & = & (x_i[0], \ldots, x_i[m-1])
\oplus~(b_i^0, \ldots, b_i^{m-1}) \oplus (b_{i-1}^0, \ldots, b_{i-1}^{m-1})\\
& = & x_i \oplus b_i \oplus b_{i-1},
\end{eqnarray*}
where the $b_i$'s are bit-vectors defined as $b_i := (b_i^0, \dots, b_i^{n-1})$.
The PTS then prints the first $n$
bits of each of the segments $u_0, \ldots , u_{m-1}$
by going through a sequence of announcement transitions via actions
$\{a_i^k : 0\le i < n, 0 \le k < m \}$
such that $a_i^k$ is $\one$ if $u_k[i] = 1$,
and $a_i^k$ is $\zero$ if $u_k[i] = 0$.
Hence, the announcement made by participant $p_i$ is
$a_i := (a_i^0, \ldots, a_i^{m-1}) = x_i \oplus b_i \oplus b_{i-1}$
for $i \in \NatInt{0}{n-1}$.

As before, we consider the case where the first participant is the observer.
The maximal traces of the PTS in this case are
in form of $t_0 \cdots t_{n-1}\ t'$,
where 
$t_0, t_{n-1} \in \{\head, \tail\}^*\reset$,
$t_i \in \toss^*\reset$
for $i \in \NatInt{1}{n-2}$, and $t' \in \{\zero,\one\}^*$.
To prove anonymity, we define a reference system
such that the initial configurations and the actions are
the same as those of the original PTS,
except that for each $k \in \NatInt{0}{m-1}$,
the announcement bits $a_0^k, \ldots, a_{n-1}^k$ 
encoded in the maximal trace from an initial configuration $w$
are uniformly distributed over 
$\{ (a_0^k, \ldots, a_{n-1}^k) :
a_0^k \oplus \cdots \oplus  a_{n-1}^k = w_k[0] \oplus \cdots \oplus w_k[n-1],
~a_0^k = w_k[0] \oplus b_0^k \oplus b_{n-1}^k \}$.
However, note that
\[
  a_0^k \oplus \cdots \oplus  a_{n-1}^k = (a_0 \oplus \cdots \oplus  a_{n-1}) [k],
\]
and
\[
  w_k[0] \oplus \cdots \oplus w_k[n-1] = (x_0 \oplus \cdots \oplus x_{n-1})[k].
\]
It follows that the distribution of the announcements $a_i$'s
is independent of the initial configuration once the values of
$x_0 \oplus \cdots \oplus x_{n-1}$,
$x_0$, $b_0$, and $b_{n-1}$ (i.e. the information observed by the first participant) are fixed.
We then compute a probabilistic bisimulation over the disjoint union
of the original system and the reference system,
establishing the anonymity property that the first participant
cannot infer the secret bits of the other participants
from the information revealed to her.

\paragraph{Remark.}
    We have specified that an initial configuration
    $w := w_0\texttt{\#}\dots\texttt{\#}w_{m-1} \in \mathcal I$
    should encode $N(w) = \min_{0 \le i < m} |w_i|$
    secrets.
    Limited by the expressiveness of regular relations, however,
    it is impossible for a PTS to determine the value of $N(w)$
    in a fixed number of steps.
    Instead, we simulate the effect of $N(w)$ as follows:
    we mark the bits $w_0[i], \ldots, w_{m-1}[i]$
    in the $i$-th round of tosses for each $i \in \NatInt{0}{n-1}$,
    and allow the system to enter the announcing stage
    only if all bits of the segments $w_0, \ldots , w_{m-1}$ are marked.
    As a consequence, an initial configuration 
    $w := w_0\texttt{\#}\dots\texttt{\#}w_{m-1}$
    has a maximal trace containing the announcement actions
    only if $|w_k| = N(w)$ for $k \in \NatInt{0}{m-1}$.
    The PTS designed in this way is an over-approximation
    of the original system in the sense that
    the anonymity property of the former would imply that of the latter.

\subsection{Modelling the grades protocol}
\label{sec:grades}

Given parameters $n\ge 3$ and $m\ge 1$,
consider a ring of $n$ participants $p_0, \ldots, p_{n-1}$ and a bound $M = 2^m$
such that each participant $p_i$ possesses a secret $x_i \in \{0,\dots, M-1\}$.
The protocol consists of two stages.
At stage one, each pair of neighboring participants $p_i$ and $p_{i+1}$
computes a random integer $u_i \in \{0,\dots, M-1\}$ by tossing a coin for $m$ times.
At stage two, each participant $p_i$ announces
$a_i := \mod{(x_i + u_i - u_{i-1})}{M}$
to the other participants. It is easy to see that
$\mod{a_0 + \cdots + a_{n-1}}{M} = \mod{x_0 + \cdots + x_{n-1}}{M}$.
Particularly,
when there exists an integer $g$ such that $(g-1) \cdot n < M$ and $x_0, \dots, x_{n-1} < g$,
it holds that $\mod{a_0 + \cdots + a_{n-1}}{M} = x_0 + \cdots + x_{n-1}$.

We model the grades protocol as follows.
An initial configuration $w$
consists of $n$ bit-vectors $w_0, \dots, w_{n-1}$ such that
$w_i := ((b_i^0, c_i^0), \dots , (b_i^{m-1}, c_i^{m-1})) \in \{0,1\}^{2m}$.
We use $(b_i^0, \dots, b_i^{m-1})$ to encode $x_i$
for $i \in \NatInt{0}{n-1}$.
We say that the configuration $w$ is \emph{compatible}
with $n$ integers $u_0, \dots, u_{n-1} \in \{0,\dots, M-1\}$
if for $k \in \NatInt{0}{m-1}$ and $i \in \NatInt{0}{n-1}$,
$c_i^{k}$ is the parity of the $k$-th carry in $\mod{(x_i + u_i - u_{i-1})}{M}$.
If $w$ is compatible with $u_0, \dots, u_{n-1}$, then
it holds that
\[
    (c_i^0, \ldots, c_i^{m-1}) \oplus x_i \oplus u_i \oplus u_{i-1} =  \mod{(x_i + u_i - u_{i-1})}{M}.
\]
Conversely, given $x_i$ and $u_i$ for $i \in \NatInt{0}{n-1}$,
there exists a unique initial configuration $w$
compatible with $u_0, \dots, u_{n-1}$.
We therefore employ a transition system similar to that
of the generalized dining cryptographers protocol
(cf.~Appendix~\ref{sec:gDCP})
to produce announcements $a_0, \ldots, a_{n-1}$.
We over-approximate the behaviors of the protocol by allowing
an initial configuration to produce announcements
based on both compatible and incompatible random integers.

To prove anonymity, we define a reference system
of which the initial configurations and the actions
are the same as those of the original system.
The transitions are specified such that
for each $k \in \NatInt{0}{m-1}$,
the announcement bits $a_0^k, \ldots, a_{n-1}^k$ 
encoded in the maximal trace from
an initial configuration
are uniformly distributed over 
$\{ (a_0^k, \ldots, a_{n-1}^k) :
a_0^k \oplus \cdots \oplus  a_{n-1}^k =
b_0^k \oplus \cdots \oplus b_{n-1}^k
\oplus c_0^k \oplus \cdots \oplus c_{n-1}^k,
~a_0^k = b_0^k \oplus c_0^k \oplus u_0^k \oplus u_{n-1}^k \}$.
Note that
\[
    a_0^k \oplus \cdots \oplus  a_{n-1}^k = b_0^k \oplus \cdots \oplus b_{n-1}^k
    \oplus c_0^k \oplus \cdots \oplus c_{n-1}^k
\]
implies that
\[
  (\mod{a_0+\cdots+a_{n-1}}{M})[k]=(\mod{x_0+\cdots+x_{n-1}}{M})[k].
\]
The distribution of the announcements $a_i$'s
is thus independent of a specific initial configuration once
the values of $\mod{x_0 + \cdots + x_{n-1}}{M}$,
$c_0,\ldots, c_{n-1}$,
$x_0$, $u_0$, and $u_{n-1}$ are fixed.
However, if we compute a probabilistic bisimulation
$R\supseteq \{ (w,w') \in \mathcal I \times \mathcal I'  :  w = w' \}$
as before, we would only verify
indistinguishability between initial configurations with
the same parity $c_0,\ldots, c_{n-1}$ of carry bits.
Instead, we strengthen our proof
by computing a probabilistic bisimulation
$R \supseteq \{(w, w') \in \mathcal I \times \mathcal I' :
\exists n.~N(w) = N(w') =n \wedge \forall i\in \NatInt{0}{n-1}.~x_i = x_i'\}$,
where $N(w)$ denotes the number of participants encoded in configuration $w$.
This establishes the desired anonymity property
that the first participant cannot infer the secrets
of the other participants from the information revealed to her in the protocol.

\subsection{A note on the modelling of the case studies}

Our tool (see Section~\ref{sec:learning}) allows the user
to refine the proof space by specifying a ``quotienting'' function
$q: \Sigma^* \to \Sigma^*$ over terminal configurations.
More precisely, the tool only explores witness bisimulations $R$
satisfying the following condition:
if $u$ and $v$ are terminal configurations, then
$(u,v) \in R$ iff $q(u) = q(v)$.
For example, $q(u) = \varepsilon$ for all $u \in \Sigma^*$ specifies that
all terminal configurations should be equivalent,
while $q(u) = u$ for all $u \in \Sigma^*$ means that a terminal configuration
is equivalent only to itself.
This design enables more flexibility for the modelling and verification
of a case study. For example,
\begin{itemize}
    \item In all of the three protocols we consider,
    the participants are required to make some information public at the end.
    This fact may be modelled by specifying that
    a maximal trace from a valid initial configuration
    should ends with a suffix encoding the information,
    e.g., an announcement $a \in (\one + \zero)^*$.
    Alternatively, we can choose to record this information in the configuration,
    e.g., by specifying that a maximal trace should end at a configuration
    $f(a)$, where $f : (\one + \zero)^* \to \Sigma^*$ is injective,
    and consider the witness bisimulations that equate a terminal configuration only to itself.
    \item 
    In the generalized dining cryptographers and the grades protocol,
    the set of valid initial configurations is not regular.
    (Recall that the set of valid initial configurations
    in the generalized dining cryptographers protocol is
    $\{ w_0\texttt{\#}\dots\texttt{\#}w_{m-1} \in \mathcal I : |w_i| = |w_j|\ \mbox{for}\ i,j\in \NatInt{0}{m-1}\}$.)
    We have circumvented this issue by over-approximating the system
    with a regular initial set $\mathcal I$
    containing both valid and invalid initial configurations.
    This approximation, however, introduces spurious terminal configurations
    that are reachable only from invalid initial configurations.
    By specifying the quotienting function, we can
    choose whether or not the spurious terminal configurations
    should be equivalent to each other.
\end{itemize}
These kinds of modelling decisions do not affect the semantics and
the correctness of a model;
however, they do change the shape (sometimes even the existence) of
the witness bisimulations required to verify the model.
In our experiment, the learning procedure in Section~\ref{sec:learning}
finds a proof for the generalized dining cryptographers protocol
only when all spurious terminal configurations are regarded as equivalent.
In contrast, the procedure finds a proof for the grades protocol
only when distinct spurious terminal configurations are regarded as non-equivalent.
Generally, it is not easy to characterize the modelling decisions
(e.g.~the choice of the quotienting function)
that could benefit the convergence of the learning procedure;
our suggestion for now is to approach the right decisions through a process of trial and error.